\documentclass[aps,prd,reprint,showpacs,floatfix,nofootinbib,superscriptaddress,byrevtex]{revtex4-1}
\usepackage{amsmath}
\usepackage{amssymb}
\usepackage{latexsym}
\usepackage{graphics}
\usepackage{graphicx}
\usepackage{slashed}
\usepackage{color}
\DeclareGraphicsExtensions{.eps,.png}

\usepackage{bm}
\usepackage[bookmarks=true,bookmarksopen=true,bookmarksnumbered=true,bookmarksopenlevel=3]{hyperref}

\begin{document}

\title{Doubly charged heavy leptons at LHC via contact interactions}

\author{\textsc{R.~Leonardi}} 
\affiliation{Dipartimento di Fisica, Universit\`{a} degli Studi di Perugia, Via A.~Pascoli, I-06123, Perugia, Italy}
\affiliation{Istituto Nazionale di Fisica Nucleare, Sezione di Perugia, Via A.~Pascoli, I-06123 Perugia, Italy}

\author{\textsc{O.~Panella}}
\affiliation{Istituto Nazionale di Fisica Nucleare, Sezione di Perugia, Via A.~Pascoli, I-06123 Perugia, Italy}
\author{\textsc{L.~Fan\`{o}}}
\affiliation{Dipartimento di Fisica, Universit\`{a} degli Studi di Perugia, Via A.~Pascoli, I-06123, Perugia, Italy}
\affiliation{Istituto Nazionale di Fisica Nucleare, Sezione di Perugia, Via A.~Pascoli, I-06123 Perugia, Italy}

\date{\today}

\begin{abstract}
We study the production of  doubly charged excited leptons at the LHC. These exotic states are predicted in extended weak isospin composite models. A recent analysis of such exotic states was based on a pure gauge model with magnetic type interactions. We include here the mechanism of contact interactions and  show that this turns out to  dominate the production of the doubly charged leptons. We perform a feasibility analysis of the observation of the tri-lepton signature associated with the production of the exotic doubly charged lepton simulating the response of a generic detector. We give exclusion plots in the parameter space, within statistical uncertainties, at different luminosities.

\end{abstract}


\maketitle

\section{Introduction}
\label{sec_intro}

Notwithstanding the recent experimental observation of the Higgs boson~\cite{Chatrchyan:2012aa,*Aad:2012aa} at the large hadron collider (LHC) at CERN, which is only the last of a spectacular series of experimental observations in total agreement with the highly successful Standard Model (SM), the theory of the fundamental interactions of elementary particles,  an impressive effort is at the same time being undertaken by the high energy physics community to unravel signals of possible scenarios beyond the Standard Model (BSM).  The recent runs at $\sqrt{s}=7,8$ TeV of the LHC accelerator  have witnessed a tremendous amount of new analyses that are continuously pushing up exclusion bounds on the parameter spaces of many BSM models.

Compositeness of ordinary fermions is one possible scenario of new physics which is receiving quite lot of attention both from the experimental and theoretical point of view.  It has been realised quite a while back~\cite{Eichten:1983hw} that if leptons and quarks do have a further level of substructure one may generically expect both $(i)$ \emph{excited states}, generally referred to as excited quarks and  leptons, and $(ii)$ \emph{contact interactions} not only  between leptons and/or quarks but also between excited states and ordinary fermions.

Phenomenological models of excited leptons and quarks have been proposed that rely on the iso-spin invariance and through this symmetry magnetic type interactions between excited and ordinary fermions are expected~\cite{Cabibbo:1983bk,Baur:1989kv}.  

Contact interactions have also been discussed early on in connection with high energy collider physics~\cite{Eichten:1983hw,Peskin:1985symp}. 
The vast majority of the literature has concentrated with flavour conserving and flavour diagonal contact interactions. Flavour changing contact interactions are of course severely constrained by processes like $K_L^0 \to \mu e$, $D^0 \bar{D}^0$ and $K^0 \bar{K}^0$ mixing~\cite{Eichten:1983hw,Eichten:1980aa} with bounds on the compositeness scale ($\Lambda$) ranging in the interval $30 \div 800$ TeV. Nothing prevents however  to consider flavour conserving but \emph{non-diagonal} contact interactions. This is a somewhat model dependent hypothesis although it suffice to assume that  different flavour quarks and/or leptons share the same constituents, which surely it is not a very far-fetched assumption.

From the experimental point of view a substantial amount of literature is available  regarding searches at high energy colliders reporting both about  the direct search of excited states (quarks and leptons), and the study of indirect effects of  four fermion contact interactions.  
Currently the stronger bounds are provided from the LHC experiments~\cite{Nakamura:2010zzi,Khachatryan:2010jd,Abe:1994aa}. 
A recent analysis of the photon + jet final state at $\sqrt{s}= 8$ TeV and with data corresponding to an integrated luminosity of 19.7 fb$^{-1}$ the CMS experiment has excluded excited quarks with masses ($m_*$) up to 3.5 TeV~\cite{CMS-PAS-EXO-13-003}. 

In a previous work \cite{Biondini:2012ny} the production of the doubly charged excited leptons $L^{++}$ within an extended isospin model based purely on gauge interactions~\cite{Pancheri:1984sm} (no contact interactions) was studied with regards to possible signals at  the LHC.
Less  studied are however the flavour non-diagonal contact interactions which could equally produce the exotic doubly charged excited leptons.  Indeed the first experimental search for contact interactions in the charged current channel ($q \bar{q}' \to e \nu$) is from the CDF experiment~\cite{Affolder:2001aa}.  
It would be therefore interesting to consider also the contribution of possible flavor conserving~\cite{Baur:1989kv} but \emph{non-diagonal} contact interactions~\cite{Eichten:1984eu,Abolins:1983} which  could trigger  the single production of the exotic doubly charged leptons  ($u\bar{d} \to L^{++}\ell^-$) object of this study. These  interactions rely on the further model dependent assumption that  also quarks and leptons of different flavour would have to  share common constituents, (in addition to quarks and leptons). 
While in the previous work \cite{Biondini:2012ny} the attention was focused on the pure gauge model with magnetic type transition couplings here we present a detailed analysis of the interplay between flavor non-diagonal contact interactions and the pure gauge model. The contact interaction is implemented in the CalcHEP (pure gauge) model developed in \cite{Biondini:2012ny} in order to study the interplay of the two mechanisms. It will be shown that the contact interaction mechanism dominates the production process while the decay is dominated by the gauge interactions in large portions of the parameter space ($\Lambda, m_*$) as long as  $m_* \lessapprox  \Lambda/3$.

We also recall that doubly charged fermions are predicted in a variety of beyond the standard model scenarios: a) extra dimensional models including custodial taus~\cite{Csaki:2008aa,Chen:2010aa,*delAguila:2010es,Kadosh:2010aa,Aguila:2010aa,Delgado:2011iz}; b) in string inspired models~\cite{Cvetic:2011iq}; c) in supersymmetric extensions of left-right symmetric models, where they appear as the supersymmetric partners of the doubly charged scalar fields~\cite{Demir:2008wt,Frank:2007nv,Babu:2013aa,Franceschini:2013aa}. Doubly charged leptons have also been considered in generic lepton triplets minimally coupling to the standard model fields~\cite{Chua:2010me}. For a model independent bottom-up approach which privileges only the charge quantum number of the exotic particles and accounts only for the simplest representations of the $SU(2)_L$ group (singlet, doublet and triplet) we refer the reader to ref.~\cite{Alloul:2013raa}.

We finally remark that the ATLAS and CMS collaborations have recently reported mass exclusion lower li\-mits for multi-charged (and long lived) particles. Based on the run at $\sqrt{s}=7$ TeV with an integrated luminosity of $L=4.4$ fb$^{-1}$ the ATLAS collaboration excludes  doubly charged particles with  masses from 50 GeV up to 430 GeV~\cite{Aad:2013aa}. The CMS Collaboration has published a combined analysis of the run at $\sqrt{s}=7$ TeV with $L=5$ fb$^{-1}$, which excludes doubly charged particles up to $\approx 550$ GeV, and the run at $\sqrt{s}=8$ TeV with $L=18.8$ fb$^{-1}$ where doubly charged particles are excluded up to $685$ GeV~\cite{Chatrchyan:2013ab}. These bounds apply to long lived particles (which decay outside the detector) and are based on the assumption of Drell-Yan-like  pair production, and do not apply, strictly speaking, to promptly decaying doubly charged particles like those that we consider here. 

The rest of the paper is organised as follows: Section~\ref{sec_model} describes briefly the details of the model and the implementation in the CalcHEP generator. 
Section~\ref{sec_production} will discuss the decay and production of the exotic doubly charged lepton comparing contact and gauge interaction mechanisms. Section~\ref{sec_signal+background} focusses on the final tri-lepton signature and the corresponding SM background. Section~\ref{sec_fast+simulation} is devoted to a discussion of fast detector simulation. Finally section~\ref{sec_conclusions} gives the final discussion and the conclusions.

\section{Extended Isospin Model}\label{sec_model}

In hadronic physics the strong isospin symmetry allowed to discover the baryon and meson resonances even when quarks and gluons were still unknown. The properties of the hadronic states could be delineated using the SU(2) and SU(3) symmetries. In analogy with this we may expect that, for the electroweak sector, the weak isospin spectroscopy could reveal some properties of excited fermions without reference to an internal structure.

The standard model fermions have $I_W=0$ and $I_W=1/2$ and the electroweak bosons have $I_W=0$ and $I_W=1$, so, combining them, we can consider fermionic excited states with $I_W\leq 3/2$. The multiplets with $I_W=1$ (triplet) and $I_W=3/2$ (quadruplet) include the doubly charged leptons that are studied in this work: 
\[ L_1 = \left( \begin{array}{l}
L^{0} \\
L^{-} \\
L^{--} \end{array} \right) ,
\qquad
L_{3/2} = \left( \begin{array}{l}
L^{+} \\
L^{0} \\
L^{-} \\
L^{--} \end{array} \right)\]
With similar multiplets for the antiparticles.
While referring to the original work in \cite{Pancheri:1984sm} for a detailed discussion of all couplings and interactions we discuss  here only the main features of the higher multiplets and write down only   the relevant effective lagrangian density.  We refer to \cite{Biondini:2012ny} for further details and here we mention only  that the higher isospin multiplets ($I_W=1,3/2$) contribute only to the iso-vector current and do not contribute to the hyper-charge current. Therefore the particles of these higher multiplets interact with the standard model fermions only via the $W$ gauge field. For the exotic doubly charged lepton of the  $I_{W}=1$ triplet and the one of the $I_{W}=3/2$ quadruplet the relevant interaction lagrangians  are respectively:
\begin{subequations}
\label{lagG}
\begin{align}
\label{lagG1}
\mathcal{L}_{\text{G}}^{(1)}=\frac{gf_{1}}{\Lambda}\left( \bar{L} \, \sigma_{\mu \nu} \, \partial^{\nu} \, W^{\mu} \, \frac{1+\gamma^{5}}{2} \, \ell \right)  + h. c.\\
\label{lagG32}
\mathcal{L}_{\text{G}}^{(3/2)}=
 \frac{gf_{3}}{\Lambda}\left( \bar{L}\sigma_{\mu \nu} \, \partial^{\nu} \, W^{\mu} \, \frac{1-\gamma^{5}}{2} \, \ell  \right)  + h.c.
 \end{align}
\end{subequations}
In the above Eqs.~(\ref{lagG1},\ref{lagG32}), $L,\ell$ are respectively the excited and ordinary fermion spinor, and $f_{1}, f_{3}$ are dimension-less coupling constants, 
expected to be of order one, whose precise values could  only be fixed within a specific compositeness model.  The above interaction lagrangians were implemented in a CalcHEP model in \cite{Biondini:2012ny}. 
Let us now discuss contact interactions (CI). These can be introduced~\cite{Baur:1989kv} by an effective four-fermion lagrangian of the type:
\begin{subequations}
\label{contact}
\begin{align}
\label{Lcontact}
\mathcal{L}_{\text{CI}}&=\frac{g_\ast^2}{\Lambda^2}\frac{1}{2}j^\mu j_\mu\\
\label{Jcontact}
j_\mu&=\eta_L\bar{f_L}\gamma_\mu f_L+\eta\prime_L\bar{f^\ast_L}\gamma_\mu f^\ast_L+\eta\prime\prime_L\bar{f^\ast_L}\gamma_\mu f_L + h.c.\nonumber\\&\phantom{=} +(L\rightarrow R)
\end{align}
\end{subequations}
with $g_*^2 = 4\pi$ and the $\eta$ factors equal to unity. In this work the right-handed currents, for simplicity, will be neglected.

In \cite{Biondini:2012ny} the gauge interactions in Eqs.~(\ref{lagG}) were implemented in {CalcHEP}~\cite{Pukhov:1999,Belyaev20131729} with the help of {FeynRules}~\cite{Christensen:2008py} a {Mathematica}~\cite{math} package which allows to obtain the Feynman rules of a given quantum field theory model described by a known lagrangian. In this work we have implemented the contact interaction in Eq.~\ref{contact}  in the same CalcHEP model of ref. \cite{Biondini:2012ny}. The operator in Eq.~\ref{contact} has to be entered "by hand" in CalcHEP introducing an auxiliary gauge field~\cite{Belyaev20131729} which is exchanged by the fermion currents. Once this done the CalcHEP generator can be used to in order to  
 study of the interplay with the gauge interactions in the phenomenology of the exotic states with respect to  \textit{production cross sections} and the excited particles \textit{decays}.
 \begin{figure*}[t!]
\includegraphics[width=16.0cm]{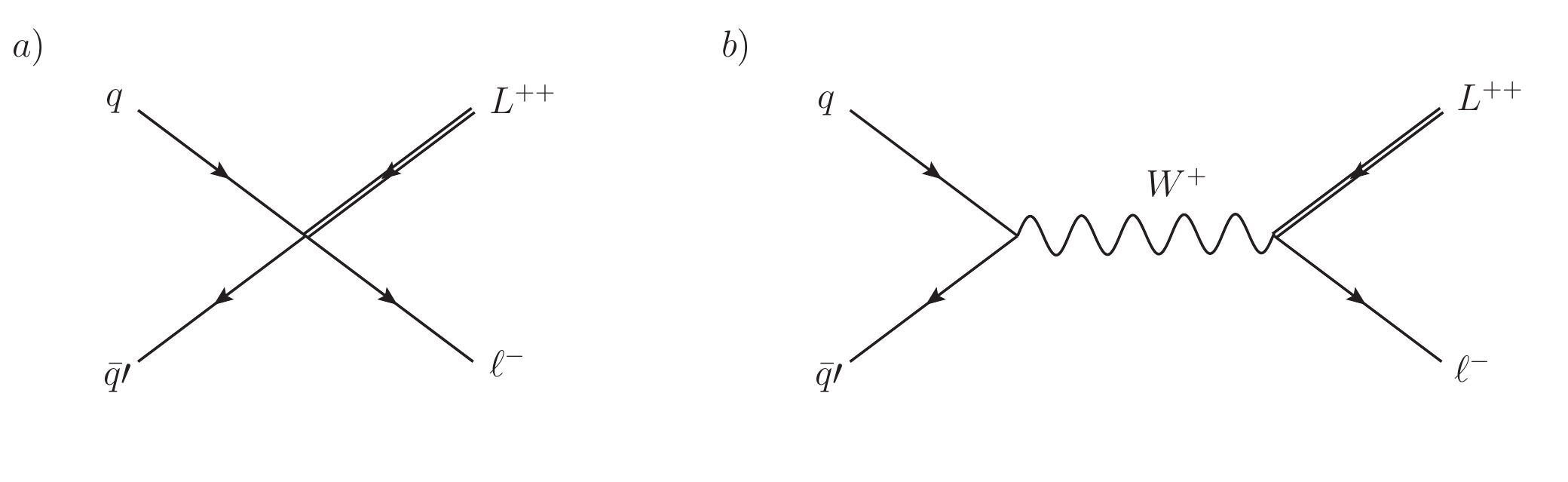}
\caption{\label{fig:fig1} The Feynman diagrams of the production processes for $L^{++}$ at the parton level $a)$ trough contact interaction $b)$ trough gauge interaction}
\end{figure*}

\section{Production and decay of the doubly charged leptons}\label{sec_production}
The doubly charged lepton is produced trough the process $pp\rightarrow L^{++} \ell^-$. At the parton level the process is $q\bar{q}'\rightarrow L^{++}\ell^-$, where $q$ is an up-type quark and $\bar{q}'$ is a down-type antiquark, and  $L^{++}$ and $\ell^-$ belong to the same lepton family,  since we assume  the conservation of lepton family number. The process can take place through both gauge and contact interaction, the two diagrams are shown in Fig.~\ref{fig:fig1}.
The parton cross section for the gauge interactions and the contact interactions are respectively:
\begin{subequations}
\begin{align}
\hat{\sigma}_{\text{GI}}&=\frac{\alpha^2}{\sin^4\theta_W}\frac{V^{qq\prime}_{CKM}}{36\pi s \Lambda^2}\frac{{\left(s-m_\ast^2\right)}^2\left(s+2m_\ast^2\right)}{{\left(s-{M_W}^2\right)}^2+{\left(M_W \Gamma_W\right)}^2 }\\
\hat{\sigma}_{\text{CI}}&=\frac{\pi}{12s}{\left[\frac{s}{\Lambda^2}\right]}^2{\left(1-\frac{m_\ast^2}{s}\right)}^2\left(1+\frac{m_\ast^2}{s}\right)\nonumber \\
& \phantom{xxxxxxxxxxxxxxxx}\times\left[1+\frac{s-m_*^2}{3(s+m_*^2)}\right]
\end{align}
\end{subequations}
where as it is customary for the contact interaction coupling constant $g_\ast$, we have used  the value $g_\ast^2=4\pi$.
\begin{figure*}[t!]
\includegraphics[width=16.0cm]{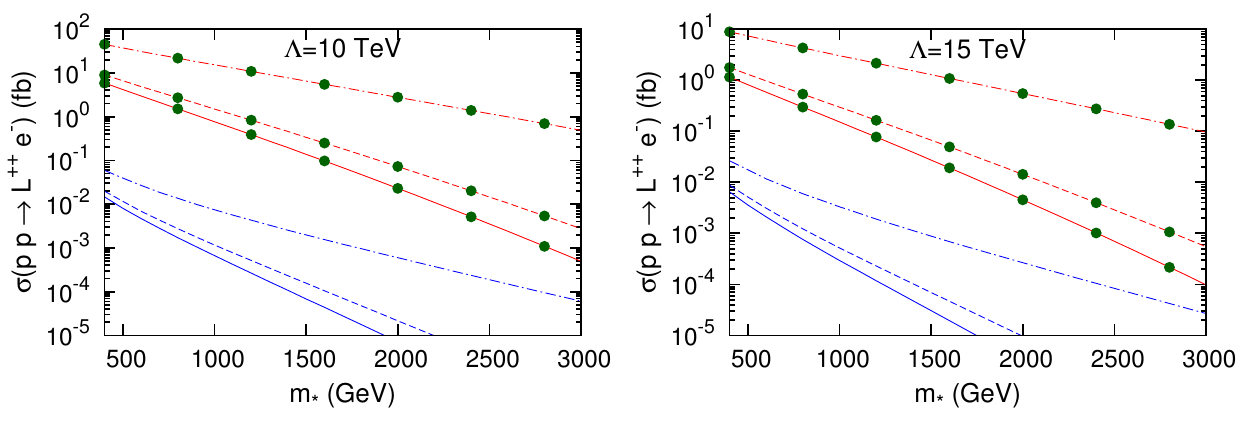}
\caption{\label{parcross} (Color on line) The production cross sections at 7 TeV (Solid line), 8 TeV (dashed line) and 14 TeV (dashed-dotted line) for contact (red on line) and gauge (blue on line) interactions; the contact interaction dominates the production. The analytical calculations are compared to the CalcHEP calculations (dots), the agreement is very good. We have used  CTEQ6m parton distribution functions~\cite{Pumplin:2002vw}.  The factorization and renormalization scale has been set to  $\hat{Q}=m_*$. The results for the gauge interactions are the same for both isospin values $I_{W}=1$ and $I_{W}=3/2$.}
\end{figure*}

We now discuss and present the production cross section for the doubly charged leptons for proton proton collisions as produced in the Cern LHC collider as stemming from the partonic collisions and cross sections $\hat{\sigma}$  discussed above. The QCD factorization theorem allows to compute the hadronic cross section in terms of convolution of the partonic cross sections $\hat{\sigma}$ and the universal parton distribution functions $f_i$ which depend on the longitudinal momenutm fractions of the partons $x$, and on the unphysical factorization scale $
\hat{Q}$:  
\begin{equation}
\sigma=\sum_{ij}\int_{\frac{m_*^2}{s}}^1 d\tau  \int_{\tau}^1 \frac{dx}{x}f_i\left(x,Q^2\right)f_j\left(\frac{\tau}{x},Q^2\right)\hat{\sigma}(\tau s,m_\ast)\nonumber
\end{equation}
where $\hat \sigma (\tau s,m_\ast)$ is the parton cross section evaluated at the scaled energy $\sqrt{\hat s}=\sqrt{\tau s}$.
We have used  CTEQ6m parton distribution functions~\cite{Pumplin:2002vw}.  The factorization and renormalization scale has been set to  $\hat{Q}=m_*$. 

In Fig.~\ref{parcross} we present the production cross section for both the gauge and the contact interaction for different values of the LHC energy, namely, $\sqrt{s}=7$,$8$,$14$ TeV.
It is evident that the contact interaction dominates the production of the doubly charged excited lepton.
In the same figure we compare the analytical results with the CalcHEP results, the agreement is very good, this ensure that the CalcHEP model has been built correctly.
We have used the CTEQ6m parton distribution functions.
The dacay of the doubly charged lepton can take place both through the gauge and contact interactions, and the possible decays are
\[ L^{++}\rightarrow \ell^+ \ell'^+\nu_{\ell'} \, ,\qquad  L^{++}\rightarrow \ell^+ q\bar{q\prime}\]
where the lepton family of $\ell'$ can be equal to or different from the family of $\ell$, $q$ is an up-type quark and $\bar{q\prime}$ is a down-type antiquark.
We computed the contribution due to the gauge and contact interactions amplitudes as well as the intereference of the two mechanisms. For the decays which involve only light  leptons and quarks, the widths are computed by stright-forward methods. In particular within the narrow-width approximation~\cite{Kauer2007413} we find: 
\begin{eqnarray}
\Gamma_{\text{light}}&=&{\left[\frac{g_\ast^2}{\Lambda^2}\right]}^2\frac{{m_\ast}^5}{1536\pi^3}N_cF_s\nonumber\\
&&+\frac{m_\ast{\left(V_{CKM}^{f_1f_2}\right)}^2}{1536\pi^3}{\left(g_G^2\right)}^2{\left(\frac{m_\ast^2}{\Lambda}\right)}^2N_c\, \frac{\pi M_W}{\Gamma_W}\times\nonumber\\
&&\phantom{xxxxxx}{\left(1-\frac{M_W^2}{{m_*}^2}\right)}^2\left(2+\frac{M_W^2}{{m_*}^2}\right)
\label{Gammalight}
\end{eqnarray}
where $N_c$ is the color factor  of the Feynman diagram and $F_S$ its symmetry factor. We note that within this approximation there is no interference effect between the gauge  and contact amplitudes. 

Next we write down an explicit expression for the three body decay widths of the exotic doubly charged lepton by allowing for one of the final state fermions to be massive in order to take properly into account the decay involving the top quark. 
For the dacays that involve the top quark, whose mass is not negligible, the width is: 
\begin{widetext}
\begin{eqnarray}
\Gamma_{\text{top}}&=&{\left[\frac{g_\ast^2}{\Lambda^2}\right]}^2\, N_c\,F_S\,\frac{m_\ast^5}{1556\,\pi^3}\,
\left[ (1-\mu^2) (1-7\mu^2-7\mu^4+\mu^6) -24\mu^4\log(\mu)\right]
\nonumber\\
&&+\frac{m_\ast{\left(V_{CKM}^{f_1f_2}\right)}^2}{3072\pi^3}{\left(g_G^2\right)}^2{\left(\frac{m_\ast^2}{\Lambda}\right)}^2 \,N_c\,
\int_{m_t^2}^{m_\ast^2}dQ^2\frac{Q^2}{{\left(Q^2-M_W^2\right)}^2+{\left(M_W\Gamma_W\right)}^2}\left(1-\frac{m_t^2}{Q^2}\right)\left(2-\frac{m_t^2}{Q^2}-\frac{m_t^4}{Q^4}\right)\times\nonumber\\&&{\left(1-\frac{Q^2}{m_*^2}\right)}^2\left(2+\frac{Q^2}{m_*^2}\right)\nonumber\\
&&+\frac{g_\ast^2g_G^2}{\Lambda^3}\frac{m_\ast^2\,V_{CKM}^{f_1f_2}}{256\sqrt{2}\pi^3}N_c\int_{m_t^2}^{m_\ast^2}dQ^2\frac{Q^2\left(Q^2-M_W^2\right)}{{\left(Q^2-M_W^2\right)}^2+{\left(M_W\Gamma_W\right)}^2}{\left(1-\frac{Q^2}{{m_*^2}}\right)}^2\left(1-\frac{m_t^2}{Q^2}\right)\left(2-\frac{m_t^2}{Q^2}-\frac{m_t^4}{Q^4}\right)
\label{Gammatop}
\end{eqnarray}
where $\mu= m_t/m_*$.
Expressions for the decay of $L^{++}$ to three light  fermions are readily found from the above by letting $m_t\to 0$. Note also that from the above Eq.~\eqref{Gammatop} the formula in Eq.~\eqref{Gammalight} is easily obtained in the limit $\Gamma_W \to 0$.
 \end{widetext}
 \begin{figure*}[t!]
\includegraphics[width=16.0cm]{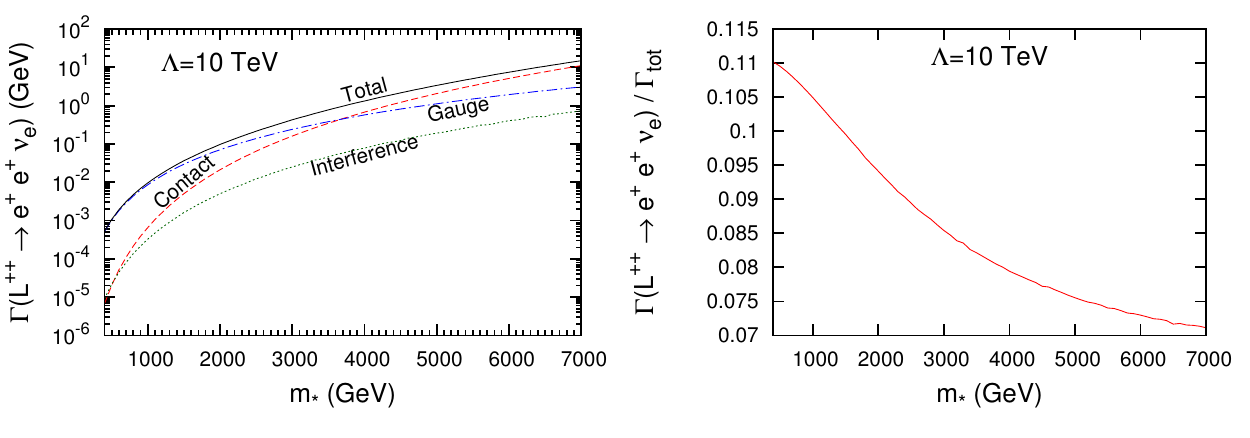}
\caption{\label{width} (Color on line) Width of the doubly charged excited letpon. One can see that over most of the parameter space analysed the decay is dominated by the gauge interactions. Only when the mass of the excited doubly charged lepton approaches the compositeness scale ($m_* \sim \Lambda$) we notice that the contact interaction mechanism becomes sizeable with the gauge mechanism and for even larger masse will eventually take over.}
\end{figure*}

\begin{figure*}[t]
\includegraphics[width=16.0cm]{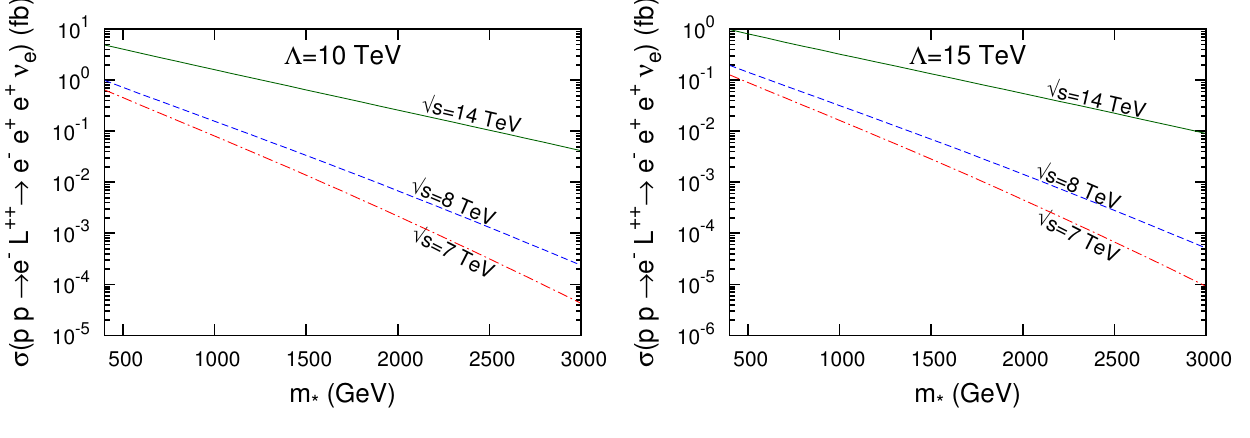}
\caption{\label{sigBr} (Color on line) The cross sections of the tri-lepton final state signature  ($pp\rightarrow \ell^-L^{++}\rightarrow \ell^-\ell^+\ell^+\nu_\ell$) at 14 TeV (solid line), 8 TeV (dashed line) and 7 TeV (dashed-dotted line). }
\end{figure*}
In Fig.~\ref{width} we compare the three contributions and their sum for the decay $L^{++}\rightarrow \ell^+\ell^+\nu_\ell$.
For each value of $\Lambda$ there is a value of the mass of the doubly charged excited lepton below which the gauge interaction dominates, while above this value the contact interaction will dominate. For $\Lambda=(10,15)$ TeV, this mass value is $\approx (3500, 5500)$ GeV.


We will now discuss the final state signature generated by the decay of the doubly charged lepton to a pair of same sign leptons and a  neutrino:
\begin{equation}
pp\rightarrow \ell^-L^{++}\rightarrow \ell^-\ell^+\ell^+\nu_\ell
\label{signature}
\end{equation}
the so-called tri-lepton final state signature which has been shown to be interesting enough even in the case of gauge interactions only~\cite{Biondini:2012ny} where it has been shown that it is characterized by a same sign dilepton   invariant mass distribution which is highly correlated with doubly charged lepton mass. 
The cross section for this signature is estimated in the small width approximation as the  production rate times the branching ratio (${\cal B}$) of the decay $L^{++}\rightarrow \ell^+\ell^+\nu_\ell$:
\begin{eqnarray}
\sigma(pp\rightarrow \ell^-\ell^+\ell^+\nu_\ell)&=&\sigma(pp\rightarrow \ell^-L^{++})\,\times\nonumber \\&&\phantom{xxxx} {\cal B}(L^{++}\rightarrow \ell^+\ell^+\nu_\ell)
\end{eqnarray}
In Fig.~\ref{sigBr} we show the results for $\sqrt{s}=7$, $8$ and $14$ TeV and for $\Lambda=10$ and $15$ TeV. We see that the cross-section values for the considered final state ($\sigma {\cal B}$) are quite encouraging. For example  at the next run with $\sqrt{s}=14$ TeV it still of the order of $4\times 10^{-2}$ fb for $\Lambda= 10$ TeV with masses as large as $m_*=3$ TeV which would provide ${\cal O}(10)$  or ${\cal O}(100)$ events  with an integrated luminosity of respectively $L=300$ fb$^{-1}$ or $L=3000$ fb$^{-1}$ (more at lower masses). These results prompted us to make a detailed analysis of the signal with respect to the standard model background in conjunction with a fast simulation of the detector response.

\section{Signal and background}
\label{sec_signal+background}
The tri-lepton final state signature considered in this work, see Eq.~\eqref{signature}
receives  background from the following standard model  processes:
\begin{align*}
pp\,\rightarrow &\,W^+Z^0\rightarrow \ell^-\ell^+\ell^+\nu_\ell \\
pp\,\rightarrow &\,W^+\gamma\rightarrow \ell^-\ell^+\ell^+\nu_\ell \\
pp\,\rightarrow &\,\ell^+\gamma^\ast\nu_\ell\rightarrow \ell^-\ell^+\ell^+\nu_\ell 
\end{align*}
This background has been studied in the detail in ref.~\cite{Biondini:2012ny} in the context of the same signature associated to the pure gauge interaction model and hence the main features can be taken from there~\cite{Biondini:2012ny}. It turns out that the background is dominated by the  $WZ$ process.

\begin{figure*}[ht!]
\begin{center}
\includegraphics[width=16.0cm]{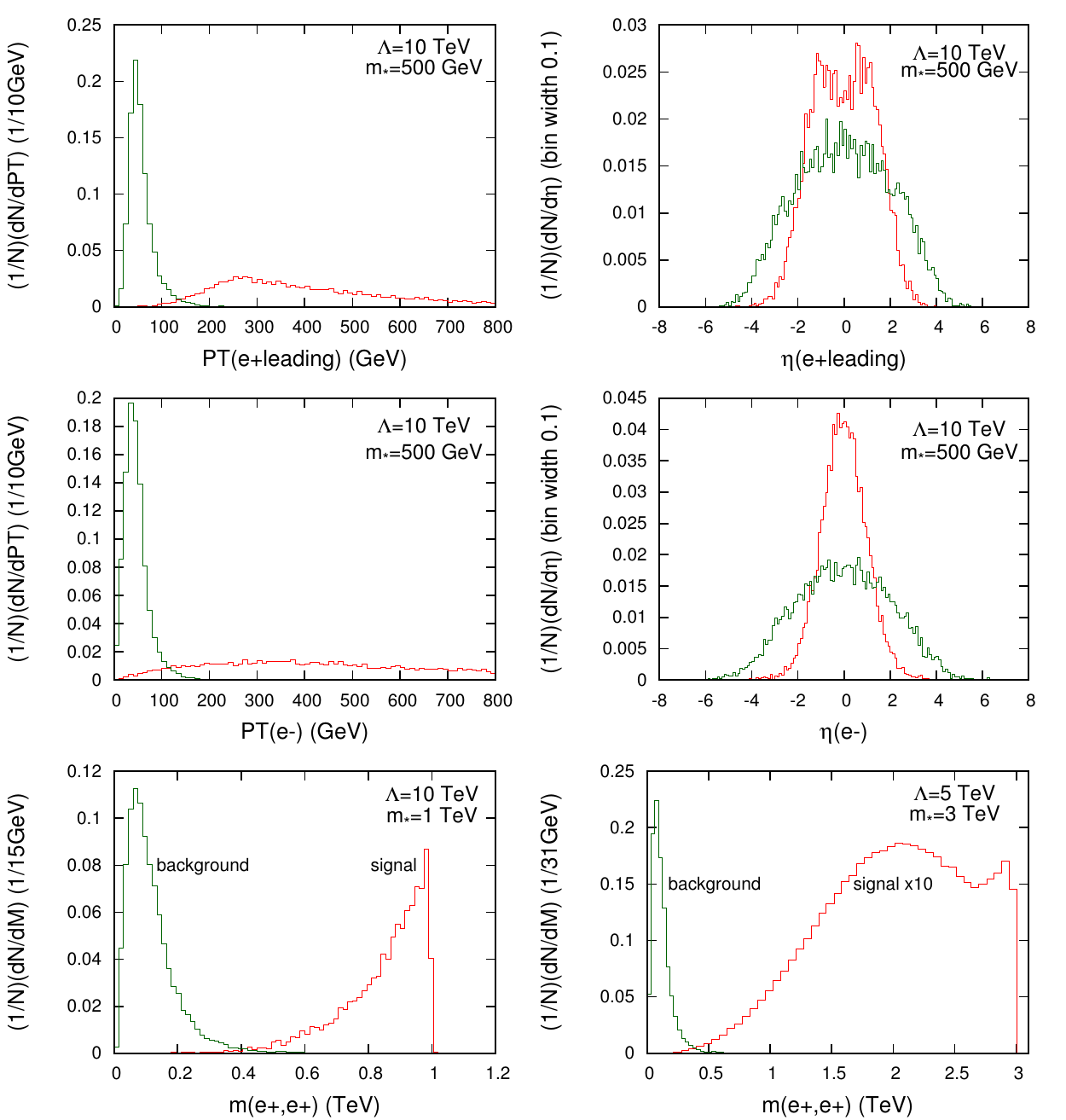}
\end{center}
\caption{\label{MISC}(Color online) Top-left: The leading positron transverse momentum distribution realized by CalcHEP. Top-right: The leading positron pseudorapidity distribution realized by CalcHEP. Center-left: The electron transverse momentum distribution realized by CalcHEP. Center-right: The electron pseudorapidity distribution realized by CalcHEP. Bottom-left: The same sign dilepton invariant mass distribution realized by CalcHEP in the region where the decay is dominated by the gauge interaction. Bottom-right: The same sign dilepton invariant mass distribution realized by CalcHEP in the region where the decay is dominated by the contact interaction.   }
\end{figure*}

We studied the differences in the kinematic distributions of signal and background by means of CalcHEP.
The transverse momentum of $\ell^+$ and $\ell^-$ in the signal is higher  than in the background, see Fig.~\ref{MISC} (Top-left, Center-left) and this is as we expect, in fact in the signal the $p_T$ of the associated lepton ($\ell^-$) has to balance the high mass of the doubly charged lepton ($L^{++}$) in order to conserve the four-momentum that initially is zero on the transverse plane and the $\ell^+$ comes from the decay of the heavy particle. In the background instead the energy scale is given by the mass of the electroweak gauge bosons, that is supposedly much lower than the excited lepton mass.
We can infer that the leading positron transverse momentum distribution allows a better discrimination between signal and background than the electron transverse momentum distribution.

By means of the study of the like-sign dilepton invariant mass distribution we can obtain information about the mass of the excited lepton. 

Indeed in the decay of $L^{++}$ the dominant interaction is the gauge one and it has the dilepton topology.
For this kind of decay, as suggested in \cite{Barr:2010zj}, the like-sign dilepton invariant mass distribution presents  a maximum very close to the end point, it is given by:
$$[m_{(\ell^+\ell^+)}^2]_{max}=\frac{(m_\ast^2-M_W^2)(M_W^2-m_{\nu}^2)}{M_W^2}\cong m_\ast^2-M_W^2 $$
The condition $m_* \gg M_W$ is generally satisfied so that $[m_{\ell^+\ell^+}]_{max}\cong m_\ast$ and the end-point of the distribution is sharply related to the mass of the exotic doubly charged lepton.

The like-sign di-lepton invariant mass distribution is shown in Fig.~\ref{MISC} (Bottom-left, Bottom-right) in comparison with that of the standard model background at a fixed point in the parameter space. The signal and background are well separated. Note that Fig.~\ref{MISC} (Bottom-left)  samples a region of the parameter space ($m_* \ll \Lambda $) where the gauge interaction dominates the decay of the doubly charged lepton and this is reflected in a dilepton invariant mass distribution which is sharply peaked  at his endpoint in line with what has been found in the pure gauge model~\cite{Biondini:2012ny}. In Fig.~\ref{MISC} (Bottom-right) we show the di-lepton invariant mass distribution for a point of the parameter space where the decay is instead dominated by the contact interactions ($m_*   \approx \Lambda$) .  We see that the distribution is much broader and it has a peak distinguished from the end point which is however always very sharp at the heavy mass $m_*$. This feature is quite important as it  might be used eventually to disentangle the pure gauge model from the one which includes contact interactions.
\section{Fast simulation and reconstructed objects}
\label{sec_fast+simulation}

CalcHEP allows to obtain kinematic distributions for particles ideally detected with an efficiency of 100\%, they aren't related reconstruction issues. In this section we want to provide a more realistic description of the signature at LHC, introducing the effects of detectors such as acceptance, efficiency and resolution in recostructing kinematic variables, this causes the spreading of the distribution and the sampling of only a part of the entire phase-space.
In order to realize this purpose we interfaced the LHE file given as output by the event generator of CalcHEP with the software DELPHES~\cite{deFavereau:2013fsa}. The LHE file contains the particles in the final state with their fourmomentum. This file can be read by DELPHES that simulates the response of a detector according to a set of parameters provided by a configuration file; a CMS-like scenario has been used.

\begin{figure*}[ht!]
\begin{center}
\includegraphics[width=16.0cm]{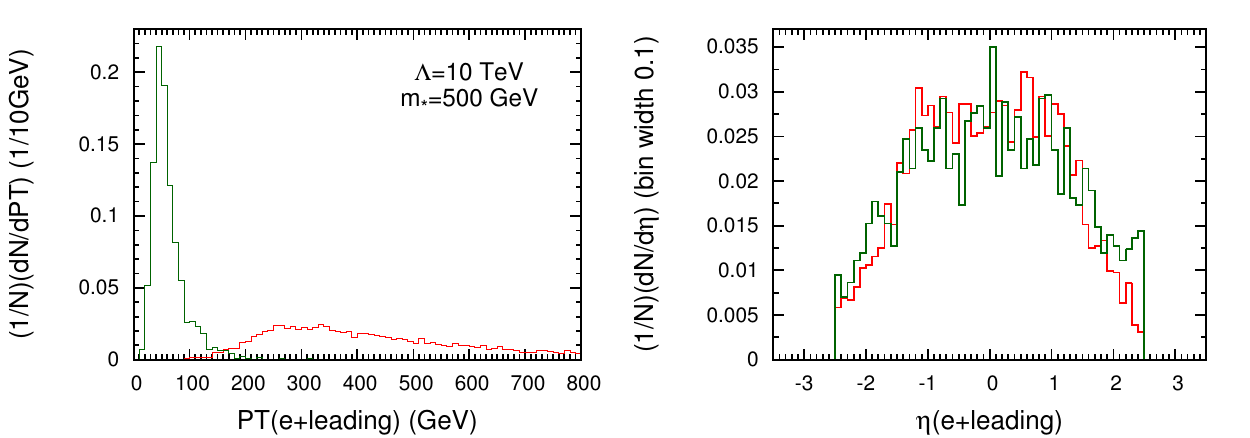}
\end{center}
\caption{\label{DIST}(Color online) Left: The leading positron transverse momentum distribution, this observable was choosen to make the kinematical cut. Right: The leading positron pseudo-rapidity distribution, for this observable the signal is not very distinct from the background.}
\end{figure*}

Firstly we considered the signal for an excited lepton of mass $m_*=500$ GeV and $\Lambda=10$ TeV and the WZ production SM background, which produces the main contribution.
For both the signal and the background we generated 10000 events, in order to have enough statistics to evaluate the efficiency of the detector. Than we selected only the events with the presence of $\ell^+\ell^+\ell^-$ uniquely reconstructed by the detector. For these events we studied the kinematic distributions to find a cut to discriminate among the signal and background.
In Fig.~\ref{DIST} (Left) we compare the reconstructed transverse momentum distribution of the leading positron both for the signal and background. The two distributions are quite separated and this clearly suggests that an high enough cut on this kinematic variable ($p_T  \gtrsim 100 $ GeV) will dramatically reduce the background while leaving substantially unaltered the signal.
In Fig.~\ref{MISC} (Top-right, Center-right) we can observe that the final state particles have a wider pseudorapidity distribution in the background relative to the signal, so we espect a lower number of accepted background events because a larger number of particles has a pseudorapidity above the detector acceptance.
Fig.~\ref{DIST} (Right) shows that pseudorapidity does not discriminate between signal and background.
Therefore the best discriminating observable is the leading positron $p_T$ and we impose the following cut:
\begin{equation}
\label{Cut}
p_T(e^+_1)>150\mbox{ }\text{GeV}
\end{equation}
where $e^+_1$ is the positron with the highest transverse momentum. Table~\ref{tabella} shows, for each selection criterion, the number of events that pass it and the corrispondent efficiency.

\begin{table*}[t!]
\caption{\label{tabella}The number of events for both signal and background after applying the selection criteria and the cut on the transverse momentum of the highest transverse momentum positron ($e_1^+$) at $\sqrt{s}=14$ TeV and with $m_*=500$ GeV.  The reconstruction efficiency and the event selection efficiency have been optimized with respect to the signal, \emph{assuming the electron channel}.}
\begin{ruledtabular}
\begin{tabular}{c|c|c|c|c}
Events & SIG ($m_*=500$ GeV) & BKG ($WZ$) & selected/total SIG & selected/total BGK\\
\hline
GENERATED EVENTS & 10000  & 10000 & 1 & 1 \\ 
\hline
Reco $e^{+} \, e^{+} \, e^{-}$& 6144 & 2428 & 0.61 & 0.24 \\
\hline
$p_{T}\left( e_1^+ \right) > 150$ GeV  & 6065 & 65 & 0.99 & 0.03\\
\end{tabular}
\end{ruledtabular}
\end{table*}

\begin{figure*}[t!]
\includegraphics[width=16.0cm]{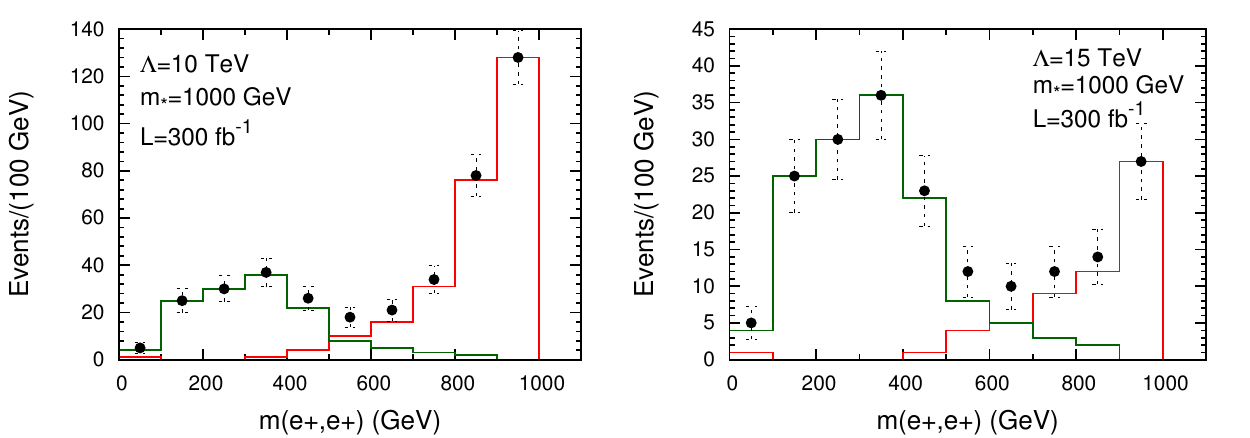}
\caption{\label{INV} (Color online) Invariant mass distribution of the same sign dilepton after the fast simulation of the detector reconstruction at $\sqrt{s}=14$ TeV and $L=300$ fb$^{-1}$. The full dots are the sum of signal and background distributions and the vertical bars show the statistical errors. We can still see the clear separation of the signal distribution, which is strongly correlated with the mass of the doubly charged lepton, from the background distribution. On the left panel we show the case ($\Lambda=10$ TeV,$m_*=1000$ TeV) while  the right panel is for  ($\Lambda=15$ TeV,$m_*=1000$ TeV)}
\end{figure*}

We present also the reconstructed  di-lepton invariant mass distribution  after the detector response considering a luminosity of 300 fb$^{-1}$, it is shown for both the signal and background in Fig.~\ref{INV}, along with an estimate of the statistical error. The signal distribution is still quite separated and visible above that of the background. This is an important conclusion which hints at this invariant mass distribution as an effective tool to pin down the doubly charged lepton mass.

\begin{figure*}[t!]
\centering
\includegraphics[width=16.0cm]{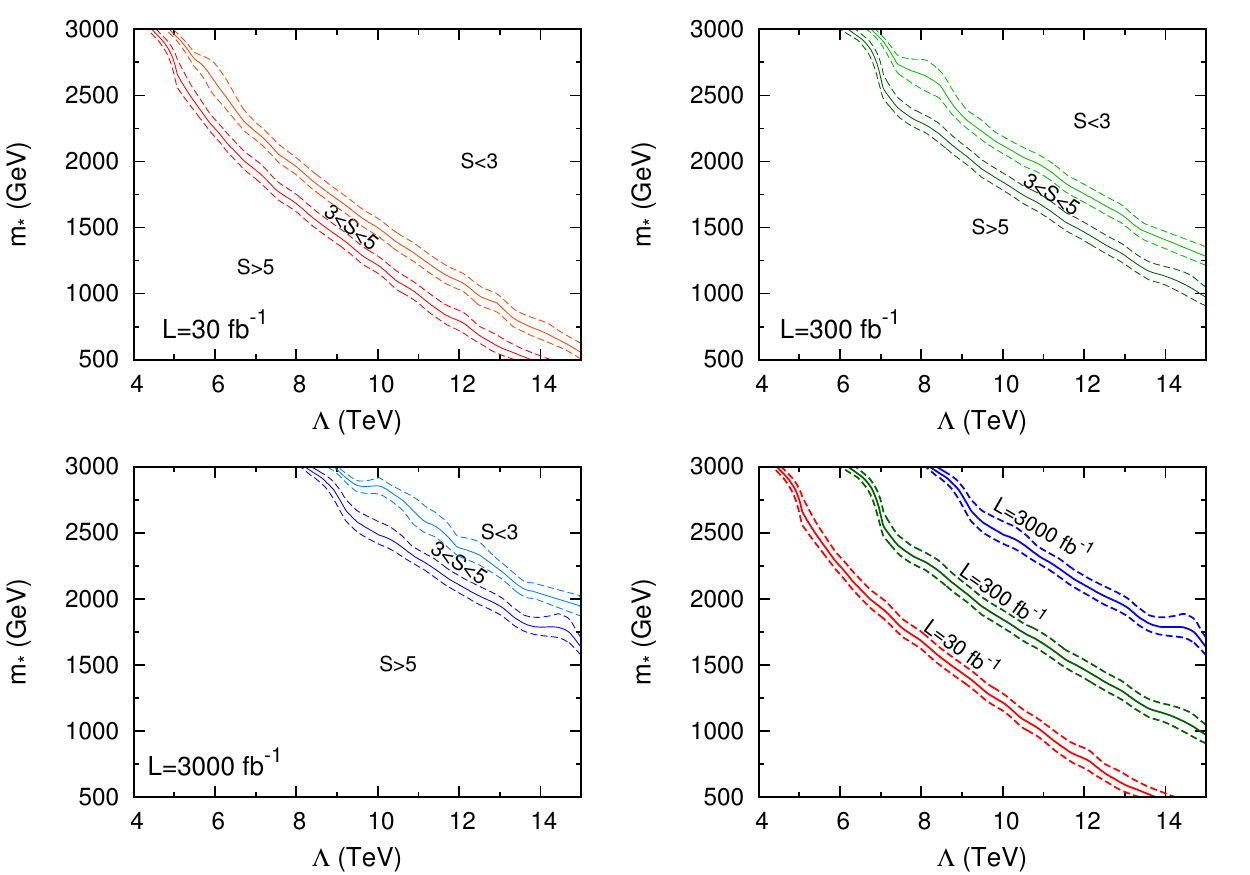}
\caption{ (Color online) Contour maps for $S=5$ and $S=3$ in the parameter space ($m_*,\Lambda$) of the statistical significance  for $\sqrt{s}=14$ TeV and for three different values of the integrated luminosity $L=30,300,3000$ fb$^{-1}$. The solid lines are the central values and the dashed lines are the extremes of the error bands. The region above the curves are excluded. In the lower right panel we compare  the 5-sigma exclusion plots at the three different values of the integrated luminosity.  \label{SS}}
\end{figure*}

Finally, at a given integrated luminosity $L$ and for any point of the parameter space ($\Lambda, m_*$) we can evaluate the expected number of events for the signal ($N_S$)  and for the  background ($N_b$) as well as the statistical significance ($S$):
\begin{equation}N_s=L\sigma_s\epsilon_s\, , \qquad N_b=L\sigma_b\epsilon_b\, , \qquad  S=\frac{N_s}{\sqrt{N_b}}\, .
\end{equation}
Applying the same procedure for a range of  values of $\Lambda$ in the interval $[4,15]$ TeV and of $m_*$ in the interval $[500,3000]$ GeV we obtained the statistical significance, within the statistical error, as function of the two parameters $(\Lambda,m_*)$ and we delimited the experimental evidence region ($S\geq 3$) and the discovery region ($S\geq 5$). We did this for three different values of luminosity $L=$ 30, 300 and 3000 fb$^{-1}$, respectively the luminosity of run I, run II and HL-LHC. The results are shown in Fig.~\ref{SS}.

\section{Discussion and conclusions}
\label{sec_conclusions}

In this work we have included contact interactions (CI) in the phenomenology of exotic composite leptons of charge $Q=+2e$ at the LHC. Such states appear in extended composite models of higher isospin multiplets, $I_{W}=1$ and $I_{W}=3/2$. In particular the doubly charged leptons $L^{++}$ of these models naturally couple with the Standard Model fermions  through the electroweak $W$-gauge boson (magnetic type gauge interactions) and their decay channels are precisely identified.  These new resonant states are expected on general grounds if a further level of substructure exists.  

In a previous paper~\cite{Biondini:2012ny} the extended isospin model, originally discussed in ref.~\cite{Pancheri:1984sm} in its essential features, has been implemented in the CalcHEP generator only with respect to the gauge interactions (GI). In this work we have included in the same CalcHEP model the contact interactions which we find to  dominate the production mechanism of the doubly charged leptons. 
The contact interaction has to be implemented \emph{by hand} in CalcHEP via an auxiliary field.  We compare some analytical results with those of CalcHEP numerical sessions, such as parton cross sections and decay widths, in order to cross-check and validate the newly defined  CalcHEP model. 
We find that in the parameter space $(\Lambda, m_*)$ the decay of the doubly charged lepton is mostly dominated by gauge interactions. More precisely for each value of $\Lambda$ there is a value of the mass of excited doubly charged lepton ($\overline{m}_*$) below which the GI dominates the decay while above $\overline{m}_*$ the decay is dominated by the CI. For $\Lambda= (10,15)$ TeV we have  $\overline{m}_* \approx (3500,5500)$ GeV.
In any case our CalcHEP model can coherently handle both contributions.

As regards the LHC phenomenology of the doubly charged leptons we concentrated on the leptonic signature deriving from the decays $ L^{++}  \rightarrow \ell^{+} \, \ell^{+} \, \nu_{\ell}$ i.e. $p\,p \rightarrow \ell^{-} \left( \ell^{+} \, \ell^{+} \right) \, \nu_{\ell}$. Thus our main interest is in the tri-lepton signature with a same sign (and same flavour) di-lepton  which is well known to be characterised by a low Standard Model background. The particular topology, allows for a clear separation of the signal and background~\cite{Biondini:2012ny}.  

We studied the main kinematic distributions for both signal and background, finding out clear differences among them. We showed that the invariant mass distribution of the same-sign dilepton system has a sharp end point corresponding to the excited lepton mass $m_{*}$. The same feature is absent in the invariant mass distribution of the SM background. Thus the invariant mass $m_{\left(\ell^{+}, \ell^{+} \right) }$ is the most discriminating variable among signal and background as shown in Fig.~\ref{MISC} (Bottom-left, Bottom-right). 



The CalcHEP generator with the new model implementing  both gauge  and contact interactions produces parton level events which refer to ideal final state particles, as an ideal detector would reveal them, i.e. detection efficiency and misidentification of particles are not considered at all. We provide a more realistic description of our processes (signal and background) developing an interface among the CalcHEP LHE output and the general purpose detector simulator DELPHES. We  apply general purpose selection cuts and identify a cut on the transverse momentum of the highest energetic lepton ($p_T(e_1)>150$ GeV) (see Eq.~\ref{Cut}) that reject background events  as much as possible (low efficiency) and save signal events  as much as possible (high efficiency). The efficiency of the detector is also considered: tracker resolution, calorimeter resolution and geometrical acceptance. The invariant mass distribution $m_{\left( {\ell^{+},\ell^{+}} \right) }$  after the loss of events due to the detector efficiency and the smearing effect are included with an estimate of the statistical error, is shown to retain its  peculiar characteristics of being ($i$) extremely well separated from that of the background and ($ii$) strongly correlated with the mass of the deputy charged lepton. See Fig.~\ref{INV}.

The final conclusion of this study is that the production of the hypothetical doubly charged leptons peculiar of extended weak isospin composite models is substantially increased by contact interactions relative to the pure gauge model studied in~\cite{Biondini:2012ny}. Preliminary fast detector simulations based on DELPHES provide the exclusion curves, with only statistical error bands, in the ($\Lambda, m_*)$ parameter space, at 3 or 5 sigma level (see Fig.~\ref{SS}) indicating mass reaches at a fixed $\Lambda =10$ TeV of $m_*\approx (1200, 1850, 2500)$ GeV with integrated luminosity of $L=(30,300,3000)$ fb$^{-1}$ respectively. 

Overall these results are quite encouraging and certainly warrant a full fledged dedicated analysis by the experimental collaborations in view of the next, upcoming, run at $\sqrt{s}=14$ TeV of the LHC accelerator.

\begin{acknowledgments}
This work is an outcome of the master thesis of R.~L. presented at the University of Perugia in September 2013.
The authors are indebted to M. Narain for having first suggested to study the phenomenology of the exotic states in the weak-isospin extended composite models.

The authors acknowledge constant interest, encouragement and support by the local Perugia CMS group.   
\end{acknowledgments}

\bibliography{biblio.bib}

\begin{thebibliography}{39}%
\makeatletter
\providecommand \@ifxundefined [1]{%
 \@ifx{#1\undefined}
}%
\providecommand \@ifnum [1]{%
 \ifnum #1\expandafter \@firstoftwo
 \else \expandafter \@secondoftwo
 \fi
}%
\providecommand \@ifx [1]{%
 \ifx #1\expandafter \@firstoftwo
 \else \expandafter \@secondoftwo
 \fi
}%
\providecommand \natexlab [1]{#1}%
\providecommand \enquote  [1]{``#1''}%
\providecommand \bibnamefont  [1]{#1}%
\providecommand \bibfnamefont [1]{#1}%
\providecommand \citenamefont [1]{#1}%
\providecommand \href@noop [0]{\@secondoftwo}%
\providecommand \href [0]{\begingroup \@sanitize@url \@href}%
\providecommand \@href[1]{\@@startlink{#1}\@@href}%
\providecommand \@@href[1]{\endgroup#1\@@endlink}%
\providecommand \@sanitize@url [0]{\catcode `\\12\catcode `\$12\catcode
  `\&12\catcode `\#12\catcode `\^12\catcode `\_12\catcode `\%12\relax}%
\providecommand \@@startlink[1]{}%
\providecommand \@@endlink[0]{}%
\providecommand \url  [0]{\begingroup\@sanitize@url \@url }%
\providecommand \@url [1]{\endgroup\@href {#1}{\urlprefix }}%
\providecommand \urlprefix  [0]{URL }%
\providecommand \Eprint [0]{\href }%
\providecommand \doibase [0]{http://dx.doi.org/}%
\providecommand \selectlanguage [0]{\@gobble}%
\providecommand \bibinfo  [0]{\@secondoftwo}%
\providecommand \bibfield  [0]{\@secondoftwo}%
\providecommand \translation [1]{[#1]}%
\providecommand \BibitemOpen [0]{}%
\providecommand \bibitemStop [0]{}%
\providecommand \bibitemNoStop [0]{.\EOS\space}%
\providecommand \EOS [0]{\spacefactor3000\relax}%
\providecommand \BibitemShut  [1]{\csname bibitem#1\endcsname}%
\let\auto@bib@innerbib\@empty
\bibitem [{\citenamefont {Chatrchyan}\ \emph {et~al.}(2012)\citenamefont
  {Chatrchyan} \emph {et~al.}}]{Chatrchyan:2012aa}%
  \BibitemOpen
  \bibfield  {author} {\bibinfo {author} {\bibfnamefont {S.}~\bibnamefont
  {Chatrchyan}} \emph {et~al.},\ }\href {\doibase
  http://dx.doi.org/10.1016/j.physletb.2012.02.064} {\bibfield  {journal}
  {\bibinfo  {journal} {Physics Letters B}\ }\textbf {\bibinfo {volume}
  {710}},\ \bibinfo {pages} {26 } (\bibinfo {year} {2012})}\BibitemShut
  {NoStop}%
\bibitem [{\citenamefont {Aad}\ \emph {et~al.}(2012)\citenamefont {Aad} \emph
  {et~al.}}]{Aad:2012aa}%
  \BibitemOpen
  \bibfield  {author} {\bibinfo {author} {\bibfnamefont {G.}~\bibnamefont
  {Aad}} \emph {et~al.},\ }\href {\doibase
  http://dx.doi.org/10.1016/j.physletb.2012.02.044} {\bibfield  {journal}
  {\bibinfo  {journal} {Physics Letters B}\ }\textbf {\bibinfo {volume}
  {710}},\ \bibinfo {pages} {49 } (\bibinfo {year} {2012})}\BibitemShut
  {NoStop}%
\bibitem [{\citenamefont {Eichten}\ \emph {et~al.}(1983)\citenamefont
  {Eichten}, \citenamefont {Lane},\ and\ \citenamefont
  {Peskin}}]{Eichten:1983hw}%
  \BibitemOpen
  \bibfield  {author} {\bibinfo {author} {\bibfnamefont {E.}~\bibnamefont
  {Eichten}}, \bibinfo {author} {\bibfnamefont {K.~D.}\ \bibnamefont {Lane}}, \
  and\ \bibinfo {author} {\bibfnamefont {M.~E.}\ \bibnamefont {Peskin}},\
  }\href {\doibase 10.1103/PhysRevLett.50.811} {\bibfield  {journal} {\bibinfo
  {journal} {Phys.Rev.Lett.}\ }\textbf {\bibinfo {volume} {50}},\ \bibinfo
  {pages} {811} (\bibinfo {year} {1983})}\BibitemShut {NoStop}%
\bibitem [{\citenamefont {Cabibbo}\ \emph {et~al.}(1984)\citenamefont
  {Cabibbo}, \citenamefont {Maiani},\ and\ \citenamefont
  {Srivastava}}]{Cabibbo:1983bk}%
  \BibitemOpen
  \bibfield  {author} {\bibinfo {author} {\bibfnamefont {N.}~\bibnamefont
  {Cabibbo}}, \bibinfo {author} {\bibfnamefont {L.}~\bibnamefont {Maiani}}, \
  and\ \bibinfo {author} {\bibfnamefont {Y.}~\bibnamefont {Srivastava}},\
  }\href {\doibase 10.1016/0370-2693(84)91850-1} {\bibfield  {journal}
  {\bibinfo  {journal} {Phys. Lett.}\ }\textbf {\bibinfo {volume} {B139}},\
  \bibinfo {pages} {459} (\bibinfo {year} {1984})}\BibitemShut {NoStop}%
\bibitem [{\citenamefont {Baur}\ \emph {et~al.}(1990)\citenamefont {Baur},
  \citenamefont {Spira},\ and\ \citenamefont {Zerwas}}]{Baur:1989kv}%
  \BibitemOpen
  \bibfield  {author} {\bibinfo {author} {\bibfnamefont {U.}~\bibnamefont
  {Baur}}, \bibinfo {author} {\bibfnamefont {M.}~\bibnamefont {Spira}}, \ and\
  \bibinfo {author} {\bibfnamefont {P.}~\bibnamefont {Zerwas}},\ }\href@noop {}
  {\bibfield  {journal} {\bibinfo  {journal} {Phys. Rev.}\ }\textbf {\bibinfo
  {volume} {D42}},\ \bibinfo {pages} {815} (\bibinfo {year}
  {1990})}\BibitemShut {NoStop}%
\bibitem [{\citenamefont {Peskin}(1985)}]{Peskin:1985symp}%
  \BibitemOpen
  \bibfield  {author} {\bibinfo {author} {\bibfnamefont {M.}~\bibnamefont
  {Peskin}},\ }\href@noop {} {\enquote {\bibinfo {title} {{International
  Symposium on Lepton Photon Interactions at High Energies}},}\ } (\bibinfo
  {year} {1985})\BibitemShut {NoStop}%
\bibitem [{\citenamefont {Eichten}\ and\ \citenamefont
  {Lane}(1980)}]{Eichten:1980aa}%
  \BibitemOpen
  \bibfield  {author} {\bibinfo {author} {\bibfnamefont {E.}~\bibnamefont
  {Eichten}}\ and\ \bibinfo {author} {\bibfnamefont {K.}~\bibnamefont {Lane}},\
  }\href {\doibase http://dx.doi.org/10.1016/0370-2693(80)90065-9} {\bibfield
  {journal} {\bibinfo  {journal} {Physics Letters B}\ }\textbf {\bibinfo
  {volume} {90}},\ \bibinfo {pages} {125 } (\bibinfo {year}
  {1980})}\BibitemShut {NoStop}%
\bibitem [{\citenamefont {Nakamura}\ \emph {et~al.}(2010)\citenamefont
  {Nakamura} \emph {et~al.}}]{Nakamura:2010zzi}%
  \BibitemOpen
  \bibfield  {author} {\bibinfo {author} {\bibfnamefont {K.}~\bibnamefont
  {Nakamura}} \emph {et~al.} (\bibinfo {collaboration} {Particle Data Group}),\
  }\href {\doibase 10.1088/0954-3899/37/7A/075021} {\bibfield  {journal}
  {\bibinfo  {journal} {J.Phys.}\ }\textbf {\bibinfo {volume} {G37}},\ \bibinfo
  {pages} {075021} (\bibinfo {year} {2010})}\BibitemShut {NoStop}%
\bibitem [{\citenamefont {Khachatryan}\ \emph {et~al.}(2010)\citenamefont
  {Khachatryan} \emph {et~al.}}]{Khachatryan:2010jd}%
  \BibitemOpen
  \bibfield  {author} {\bibinfo {author} {\bibfnamefont {V.}~\bibnamefont
  {Khachatryan}} \emph {et~al.} (\bibinfo {collaboration} {CMS
  Collaboration}),\ }\href {\doibase 10.1103/PhysRevLett.106.029902} {\bibfield
   {journal} {\bibinfo  {journal} {Phys.Rev.Lett.}\ }\textbf {\bibinfo {volume}
  {105}},\ \bibinfo {pages} {211801} (\bibinfo {year} {2010})},\ \Eprint
  {http://arxiv.org/abs/1010.0203} {arXiv:1010.0203 [hep-ex]} \BibitemShut
  {NoStop}%
\bibitem [{\citenamefont {Abe}\ \emph {et~al.}(1994)\citenamefont {Abe} \emph
  {et~al.}}]{Abe:1994aa}%
  \BibitemOpen
  \bibfield  {author} {\bibinfo {author} {\bibfnamefont {F.}~\bibnamefont
  {Abe}} \emph {et~al.} (\bibinfo {collaboration} {CDF Collaboration}),\ }\href
  {\doibase 10.1103/PhysRevLett.72.3004} {\bibfield  {journal} {\bibinfo
  {journal} {Phys. Rev. Lett.}\ }\textbf {\bibinfo {volume} {72}},\ \bibinfo
  {pages} {3004} (\bibinfo {year} {1994})}\BibitemShut {NoStop}%
\bibitem [{CMS(2014)}]{CMS-PAS-EXO-13-003}%
  \BibitemOpen
  \href@noop {} {\emph {\bibinfo {title} {{Search for excited quarks in the
  photon + jet final state in proton proton collisions at 8 TeV}}}},\ \bibinfo
  {type} {Tech. Rep.}\ \bibinfo {number} {CMS-PAS-EXO-13-003}\ (\bibinfo
  {institution} {CERN},\ \bibinfo {address} {Geneva},\ \bibinfo {year}
  {2014})\BibitemShut {NoStop}%
\bibitem [{\citenamefont {Biondini}\ \emph {et~al.}(2012)\citenamefont
  {Biondini}, \citenamefont {Panella}, \citenamefont {Pancheri}, \citenamefont
  {Srivastava},\ and\ \citenamefont {Fano}}]{Biondini:2012ny}%
  \BibitemOpen
  \bibfield  {author} {\bibinfo {author} {\bibfnamefont {S.}~\bibnamefont
  {Biondini}}, \bibinfo {author} {\bibfnamefont {O.}~\bibnamefont {Panella}},
  \bibinfo {author} {\bibfnamefont {G.}~\bibnamefont {Pancheri}}, \bibinfo
  {author} {\bibfnamefont {Y.}~\bibnamefont {Srivastava}}, \ and\ \bibinfo
  {author} {\bibfnamefont {L.}~\bibnamefont {Fano}},\ }\href {\doibase
  10.1103/PhysRevD.85.095018} {\bibfield  {journal} {\bibinfo  {journal}
  {Phys.Rev.}\ }\textbf {\bibinfo {volume} {D85}},\ \bibinfo {pages} {095018}
  (\bibinfo {year} {2012})},\ \Eprint {http://arxiv.org/abs/1201.3764}
  {arXiv:1201.3764 [hep-ph]} \BibitemShut {NoStop}%
\bibitem [{\citenamefont {Pancheri}\ and\ \citenamefont
  {Srivastava}(1984)}]{Pancheri:1984sm}%
  \BibitemOpen
  \bibfield  {author} {\bibinfo {author} {\bibfnamefont {G.}~\bibnamefont
  {Pancheri}}\ and\ \bibinfo {author} {\bibfnamefont {Y.}~\bibnamefont
  {Srivastava}},\ }\href {\doibase 10.1016/0370-2693(84)90649-X} {\bibfield
  {journal} {\bibinfo  {journal} {Phys.Lett.}\ }\textbf {\bibinfo {volume}
  {B146}},\ \bibinfo {pages} {87} (\bibinfo {year} {1984})}\BibitemShut
  {NoStop}%
\bibitem [{\citenamefont {Affolder~et. al.}(2001)}]{Affolder:2001aa}%
  \BibitemOpen
  \bibfield  {author} {\bibinfo {author} {\bibfnamefont {T.}~\bibnamefont
  {Affolder~et. al.}} (\bibinfo {collaboration} {CDF Collaboration}),\ }\href
  {\doibase 10.1103/PhysRevLett.87.231803} {\bibfield  {journal} {\bibinfo
  {journal} {Phys. Rev. Lett.}\ }\textbf {\bibinfo {volume} {87}},\ \bibinfo
  {pages} {231803} (\bibinfo {year} {2001})}\BibitemShut {NoStop}%
\bibitem [{\citenamefont {Eichten}\ \emph {et~al.}(1984)\citenamefont
  {Eichten}, \citenamefont {Hinchliffe}, \citenamefont {Lane},\ and\
  \citenamefont {Quigg}}]{Eichten:1984eu}%
  \BibitemOpen
  \bibfield  {author} {\bibinfo {author} {\bibfnamefont {E.}~\bibnamefont
  {Eichten}}, \bibinfo {author} {\bibfnamefont {I.}~\bibnamefont {Hinchliffe}},
  \bibinfo {author} {\bibfnamefont {K.~D.}\ \bibnamefont {Lane}}, \ and\
  \bibinfo {author} {\bibfnamefont {C.}~\bibnamefont {Quigg}},\ }\href
  {\doibase 10.1103/RevModPhys.56.579} {\bibfield  {journal} {\bibinfo
  {journal} {Rev.Mod.Phys.}\ }\textbf {\bibinfo {volume} {56}},\ \bibinfo
  {pages} {579} (\bibinfo {year} {1984})}\BibitemShut {NoStop}%
\bibitem [{\citenamefont {Abolins~et al.}(1983)}]{Abolins:1983}%
  \BibitemOpen
  \bibfield  {author} {\bibinfo {author} {\bibfnamefont {M.~A.}\ \bibnamefont
  {Abolins~et al.}},\ }\href@noop {} {\enquote {\bibinfo {title} {{Proceedings
  of the 1982 DPF Summer Study on Elementary Particle Physics and Future
  Facilities, Snowmass}},}\ }\bibinfo {howpublished} {edited by R. Donaldson,
  R. Gustafson and F. Paige} (\bibinfo {year} {(APS, New York, 1983)}),\
  \bibinfo {note} {p 274}\BibitemShut {NoStop}%
\bibitem [{\citenamefont {Cs{\'a}ki}\ \emph {et~al.}(2008)\citenamefont
  {Cs{\'a}ki}, \citenamefont {Delaunay}, \citenamefont {Grojean},\ and\
  \citenamefont {Grossman}}]{Csaki:2008aa}%
  \BibitemOpen
  \bibfield  {author} {\bibinfo {author} {\bibfnamefont {C.}~\bibnamefont
  {Cs{\'a}ki}}, \bibinfo {author} {\bibfnamefont {C.}~\bibnamefont {Delaunay}},
  \bibinfo {author} {\bibfnamefont {C.}~\bibnamefont {Grojean}}, \ and\
  \bibinfo {author} {\bibfnamefont {Y.}~\bibnamefont {Grossman}},\ }\href
  {http://stacks.iop.org/1126-6708/2008/i=10/a=055} {\bibfield  {journal}
  {\bibinfo  {journal} {Journal of High Energy Physics}\ }\textbf {\bibinfo
  {volume} {2008}},\ \bibinfo {pages} {055} (\bibinfo {year}
  {2008})}\BibitemShut {NoStop}%
\bibitem [{\citenamefont {Chen}\ \emph {et~al.}(2010)\citenamefont {Chen},
  \citenamefont {Mahanthappa},\ and\ \citenamefont {Yu}}]{Chen:2010aa}%
  \BibitemOpen
  \bibfield  {author} {\bibinfo {author} {\bibfnamefont {M.-C.}\ \bibnamefont
  {Chen}}, \bibinfo {author} {\bibfnamefont {K.~T.}\ \bibnamefont
  {Mahanthappa}}, \ and\ \bibinfo {author} {\bibfnamefont {F.}~\bibnamefont
  {Yu}},\ }\href {\doibase 10.1103/PhysRevD.81.036004} {\bibfield  {journal}
  {\bibinfo  {journal} {Phys. Rev. D}\ }\textbf {\bibinfo {volume} {81}},\
  \bibinfo {pages} {036004} (\bibinfo {year} {2010})}\BibitemShut {NoStop}%
\bibitem [{\citenamefont {del Aguila}\ \emph {et~al.}(2011)\citenamefont {del
  Aguila}, \citenamefont {Carmona},\ and\ \citenamefont
  {Santiago}}]{delAguila:2010es}%
  \BibitemOpen
  \bibfield  {author} {\bibinfo {author} {\bibfnamefont {F.}~\bibnamefont {del
  Aguila}}, \bibinfo {author} {\bibfnamefont {A.}~\bibnamefont {Carmona}}, \
  and\ \bibinfo {author} {\bibfnamefont {J.}~\bibnamefont {Santiago}},\ }\href
  {\doibase 10.1016/j.physletb.2010.11.054} {\bibfield  {journal} {\bibinfo
  {journal} {Phys.Lett.}\ }\textbf {\bibinfo {volume} {B695}},\ \bibinfo
  {pages} {449} (\bibinfo {year} {2011})},\ \Eprint
  {http://arxiv.org/abs/1007.4206} {arXiv:1007.4206 [hep-ph]} \BibitemShut
  {NoStop}%
\bibitem [{\citenamefont {Kadosh}\ and\ \citenamefont
  {Pallante}(2010)}]{Kadosh:2010aa}%
  \BibitemOpen
  \bibfield  {author} {\bibinfo {author} {\bibfnamefont {A.}~\bibnamefont
  {Kadosh}}\ and\ \bibinfo {author} {\bibfnamefont {E.}~\bibnamefont
  {Pallante}},\ }\href {\doibase 10.1007/JHEP08(2010)115} {\bibfield  {journal}
  {\bibinfo  {journal} {Journal of High Energy Physics}\ }\textbf {\bibinfo
  {volume} {2010}},\ \bibinfo {pages} {1} (\bibinfo {year} {2010})}\BibitemShut
  {NoStop}%
\bibitem [{\citenamefont {{\'A}guila}\ \emph {et~al.}(2010)\citenamefont
  {{\'A}guila}, \citenamefont {Carmona},\ and\ \citenamefont
  {Santiago}}]{Aguila:2010aa}%
  \BibitemOpen
  \bibfield  {author} {\bibinfo {author} {\bibfnamefont {F.}~\bibnamefont
  {{\'A}guila}}, \bibinfo {author} {\bibfnamefont {A.}~\bibnamefont {Carmona}},
  \ and\ \bibinfo {author} {\bibfnamefont {J.}~\bibnamefont {Santiago}},\
  }\href {\doibase 10.1007/JHEP08(2010)127} {\bibfield  {journal} {\bibinfo
  {journal} {Journal of High Energy Physics}\ }\textbf {\bibinfo {volume}
  {2010}},\ \bibinfo {pages} {1} (\bibinfo {year} {2010})}\BibitemShut
  {NoStop}%
\bibitem [{\citenamefont {Delgado}\ \emph {et~al.}(2011)\citenamefont
  {Delgado}, \citenamefont {Garcia~Cely}, \citenamefont {Han},\ and\
  \citenamefont {Wang}}]{Delgado:2011iz}%
  \BibitemOpen
  \bibfield  {author} {\bibinfo {author} {\bibfnamefont {A.}~\bibnamefont
  {Delgado}}, \bibinfo {author} {\bibfnamefont {C.}~\bibnamefont
  {Garcia~Cely}}, \bibinfo {author} {\bibfnamefont {T.}~\bibnamefont {Han}}, \
  and\ \bibinfo {author} {\bibfnamefont {Z.}~\bibnamefont {Wang}},\ }\href
  {\doibase 10.1103/PhysRevD.84.073007} {\bibfield  {journal} {\bibinfo
  {journal} {Phys.Rev.}\ }\textbf {\bibinfo {volume} {D84}},\ \bibinfo {pages}
  {073007} (\bibinfo {year} {2011})},\ \Eprint {http://arxiv.org/abs/1105.5417}
  {arXiv:1105.5417 [hep-ph]} \BibitemShut {NoStop}%
\bibitem [{\citenamefont {Cvetic}\ \emph {et~al.}(2011)\citenamefont {Cvetic},
  \citenamefont {Halverson},\ and\ \citenamefont {Langacker}}]{Cvetic:2011iq}%
  \BibitemOpen
  \bibfield  {author} {\bibinfo {author} {\bibfnamefont {M.}~\bibnamefont
  {Cvetic}}, \bibinfo {author} {\bibfnamefont {J.}~\bibnamefont {Halverson}}, \
  and\ \bibinfo {author} {\bibfnamefont {P.}~\bibnamefont {Langacker}},\ }\href
  {\doibase 10.1007/JHEP11(2011)058} {\bibfield  {journal} {\bibinfo  {journal}
  {JHEP}\ }\textbf {\bibinfo {volume} {1111}},\ \bibinfo {pages} {058}
  (\bibinfo {year} {2011})},\ \Eprint {http://arxiv.org/abs/1108.5187}
  {arXiv:1108.5187 [hep-ph]} \BibitemShut {NoStop}%
\bibitem [{\citenamefont {Demir}\ \emph {et~al.}(2008)\citenamefont {Demir},
  \citenamefont {Frank}, \citenamefont {Huitu}, \citenamefont {Rai},\ and\
  \citenamefont {Turan}}]{Demir:2008wt}%
  \BibitemOpen
  \bibfield  {author} {\bibinfo {author} {\bibfnamefont {D.~A.}\ \bibnamefont
  {Demir}}, \bibinfo {author} {\bibfnamefont {M.}~\bibnamefont {Frank}},
  \bibinfo {author} {\bibfnamefont {K.}~\bibnamefont {Huitu}}, \bibinfo
  {author} {\bibfnamefont {S.~K.}\ \bibnamefont {Rai}}, \ and\ \bibinfo
  {author} {\bibfnamefont {I.}~\bibnamefont {Turan}},\ }\href {\doibase
  10.1103/PhysRevD.78.035013} {\bibfield  {journal} {\bibinfo  {journal}
  {Phys.Rev.}\ }\textbf {\bibinfo {volume} {D78}},\ \bibinfo {pages} {035013}
  (\bibinfo {year} {2008})},\ \Eprint {http://arxiv.org/abs/0805.4202}
  {arXiv:0805.4202 [hep-ph]} \BibitemShut {NoStop}%
\bibitem [{\citenamefont {Frank}\ \emph {et~al.}(2008)\citenamefont {Frank},
  \citenamefont {Huitu},\ and\ \citenamefont {Rai}}]{Frank:2007nv}%
  \BibitemOpen
  \bibfield  {author} {\bibinfo {author} {\bibfnamefont {M.}~\bibnamefont
  {Frank}}, \bibinfo {author} {\bibfnamefont {K.}~\bibnamefont {Huitu}}, \ and\
  \bibinfo {author} {\bibfnamefont {S.~K.}\ \bibnamefont {Rai}},\ }\href
  {\doibase 10.1103/PhysRevD.77.015006} {\bibfield  {journal} {\bibinfo
  {journal} {Phys.Rev.}\ }\textbf {\bibinfo {volume} {D77}},\ \bibinfo {pages}
  {015006} (\bibinfo {year} {2008})},\ \Eprint {http://arxiv.org/abs/0710.2415}
  {arXiv:0710.2415 [hep-ph]} \BibitemShut {NoStop}%
\bibitem [{\citenamefont {Babu}\ \emph {et~al.}(2013)\citenamefont {Babu},
  \citenamefont {Patra},\ and\ \citenamefont {Rai}}]{Babu:2013aa}%
  \BibitemOpen
  \bibfield  {author} {\bibinfo {author} {\bibfnamefont {K.~S.}\ \bibnamefont
  {Babu}}, \bibinfo {author} {\bibfnamefont {A.}~\bibnamefont {Patra}}, \ and\
  \bibinfo {author} {\bibfnamefont {S.~K.}\ \bibnamefont {Rai}},\ }\href
  {\doibase 10.1103/PhysRevD.88.055006} {\bibfield  {journal} {\bibinfo
  {journal} {Phys. Rev. D}\ }\textbf {\bibinfo {volume} {88}},\ \bibinfo
  {pages} {055006} (\bibinfo {year} {2013})}\BibitemShut {NoStop}%
\bibitem [{\citenamefont {Franceschini}\ and\ \citenamefont
  {Mohapatra}(2013)}]{Franceschini:2013aa}%
  \BibitemOpen
  \bibfield  {author} {\bibinfo {author} {\bibfnamefont {R.}~\bibnamefont
  {Franceschini}}\ and\ \bibinfo {author} {\bibfnamefont {R.}~\bibnamefont
  {Mohapatra}},\ }\href@noop {} {\  (\bibinfo {year} {2013})},\ \Eprint
  {http://arxiv.org/abs/1306.6108} {arXiv:1306.6108 [hep-ph]} \BibitemShut
  {NoStop}%
\bibitem [{\citenamefont {Chua}\ and\ \citenamefont {Law}(2011)}]{Chua:2010me}%
  \BibitemOpen
  \bibfield  {author} {\bibinfo {author} {\bibfnamefont {C.-K.}\ \bibnamefont
  {Chua}}\ and\ \bibinfo {author} {\bibfnamefont {S.~S.}\ \bibnamefont {Law}},\
  }\href {\doibase 10.1103/PhysRevD.83.055010} {\bibfield  {journal} {\bibinfo
  {journal} {Phys.Rev.}\ }\textbf {\bibinfo {volume} {D83}},\ \bibinfo {pages}
  {055010} (\bibinfo {year} {2011})},\ \Eprint {http://arxiv.org/abs/1011.4730}
  {arXiv:1011.4730 [hep-ph]} \BibitemShut {NoStop}%
\bibitem [{\citenamefont {Alloul}\ \emph {et~al.}(2013)\citenamefont {Alloul},
  \citenamefont {Frank}, \citenamefont {Fuks},\ and\ \citenamefont
  {de~Traubenberg}}]{Alloul:2013raa}%
  \BibitemOpen
  \bibfield  {author} {\bibinfo {author} {\bibfnamefont {A.}~\bibnamefont
  {Alloul}}, \bibinfo {author} {\bibfnamefont {M.}~\bibnamefont {Frank}},
  \bibinfo {author} {\bibfnamefont {B.}~\bibnamefont {Fuks}}, \ and\ \bibinfo
  {author} {\bibfnamefont {M.~R.}\ \bibnamefont {de~Traubenberg}},\ }\href
  {\doibase 10.1103/PhysRevD.88.075004} {\bibfield  {journal} {\bibinfo
  {journal} {Phys.Rev.}\ }\textbf {\bibinfo {volume} {D88}},\ \bibinfo {pages}
  {075004} (\bibinfo {year} {2013})},\ \Eprint {http://arxiv.org/abs/1307.1711}
  {arXiv:1307.1711 [hep-ph]} \BibitemShut {NoStop}%
\bibitem [{\citenamefont {Aad}\ \emph {et~al.}(2013)\citenamefont {Aad} \emph
  {et~al.}}]{Aad:2013aa}%
  \BibitemOpen
  \bibfield  {author} {\bibinfo {author} {\bibfnamefont {G.}~\bibnamefont
  {Aad}} \emph {et~al.},\ }\href {\doibase
  http://dx.doi.org/10.1016/j.physletb.2013.04.036} {\bibfield  {journal}
  {\bibinfo  {journal} {Physics Letters B}\ }\textbf {\bibinfo {volume}
  {722}},\ \bibinfo {pages} {305 } (\bibinfo {year} {2013})}\BibitemShut
  {NoStop}%
\bibitem [{\citenamefont {Chatrchyan}\ \emph {et~al.}(2013)\citenamefont
  {Chatrchyan} \emph {et~al.}}]{Chatrchyan:2013ab}%
  \BibitemOpen
  \bibfield  {author} {\bibinfo {author} {\bibfnamefont {S.}~\bibnamefont
  {Chatrchyan}} \emph {et~al.},\ }\href {\doibase 10.1007/JHEP07(2013)122}
  {\bibfield  {journal} {\bibinfo  {journal} {Journal of High Energy Physics}\
  }\textbf {\bibinfo {volume} {07}},\ \bibinfo {pages} {122} (\bibinfo {year}
  {2013})}\BibitemShut {NoStop}%
\bibitem [{\citenamefont {Pukhov}\ \emph {et~al.}()\citenamefont {Pukhov},
  \citenamefont {Boos}, \citenamefont {Dubinin}, \citenamefont {Edneral},
  \citenamefont {Ilyin}, \citenamefont {Kovalenko}, \citenamefont {Kryukov},
  \citenamefont {Savrin}, \citenamefont {Shichanin},\ and\ \citenamefont
  {Semenov}}]{Pukhov:1999}%
  \BibitemOpen
  \bibfield  {author} {\bibinfo {author} {\bibfnamefont {A.}~\bibnamefont
  {Pukhov}}, \bibinfo {author} {\bibfnamefont {E.}~\bibnamefont {Boos}},
  \bibinfo {author} {\bibfnamefont {M.}~\bibnamefont {Dubinin}}, \bibinfo
  {author} {\bibfnamefont {V.}~\bibnamefont {Edneral}}, \bibinfo {author}
  {\bibfnamefont {V.}~\bibnamefont {Ilyin}}, \bibinfo {author} {\bibfnamefont
  {D.}~\bibnamefont {Kovalenko}}, \bibinfo {author} {\bibfnamefont
  {A.}~\bibnamefont {Kryukov}}, \bibinfo {author} {\bibfnamefont
  {V.}~\bibnamefont {Savrin}}, \bibinfo {author} {\bibfnamefont
  {S.}~\bibnamefont {Shichanin}}, \ and\ \bibinfo {author} {\bibfnamefont
  {A.}~\bibnamefont {Semenov}},\ }\href@noop {} {}\bibinfo {howpublished}
  {arXiv:hep-ph/9908288v2}\BibitemShut {NoStop}%
\bibitem [{\citenamefont {Belyaev}\ \emph {et~al.}(2013)\citenamefont
  {Belyaev}, \citenamefont {Christensen},\ and\ \citenamefont
  {Pukhov}}]{Belyaev20131729}%
  \BibitemOpen
  \bibfield  {author} {\bibinfo {author} {\bibfnamefont {A.}~\bibnamefont
  {Belyaev}}, \bibinfo {author} {\bibfnamefont {N.~D.}\ \bibnamefont
  {Christensen}}, \ and\ \bibinfo {author} {\bibfnamefont {A.}~\bibnamefont
  {Pukhov}},\ }\href {\doibase http://dx.doi.org/10.1016/j.cpc.2013.01.014}
  {\bibfield  {journal} {\bibinfo  {journal} {Computer Physics Communications}\
  }\textbf {\bibinfo {volume} {184}},\ \bibinfo {pages} {1729 } (\bibinfo
  {year} {2013})}\BibitemShut {NoStop}%
\bibitem [{\citenamefont {Christensen}\ and\ \citenamefont
  {Duhr}(2009)}]{Christensen:2008py}%
  \BibitemOpen
  \bibfield  {author} {\bibinfo {author} {\bibfnamefont {N.~D.}\ \bibnamefont
  {Christensen}}\ and\ \bibinfo {author} {\bibfnamefont {C.}~\bibnamefont
  {Duhr}},\ }\href {\doibase 10.1016/j.cpc.2009.02.018} {\bibfield  {journal}
  {\bibinfo  {journal} {Comput.Phys.Commun.}\ }\textbf {\bibinfo {volume}
  {180}},\ \bibinfo {pages} {1614} (\bibinfo {year} {2009})},\ \Eprint
  {http://arxiv.org/abs/0806.4194} {arXiv:0806.4194 [hep-ph]} \BibitemShut
  {NoStop}%
\bibitem [{mat()}]{math}%
  \BibitemOpen
  \href@noop {} {\enquote {\bibinfo {title} {{Mathematica}},}\ }\bibinfo
  {howpublished} {{(Wolfram Research, Inc., Champaign, IL, 2008)}},\ \bibinfo
  {note} {version 7.0}\BibitemShut {NoStop}%
\bibitem [{\citenamefont {Pumplin}\ \emph {et~al.}(2002)\citenamefont
  {Pumplin}, \citenamefont {Stump}, \citenamefont {Huston}, \citenamefont
  {Lai}, \citenamefont {Nadolsky} \emph {et~al.}}]{Pumplin:2002vw}%
  \BibitemOpen
  \bibfield  {author} {\bibinfo {author} {\bibfnamefont {J.}~\bibnamefont
  {Pumplin}}, \bibinfo {author} {\bibfnamefont {D.}~\bibnamefont {Stump}},
  \bibinfo {author} {\bibfnamefont {J.}~\bibnamefont {Huston}}, \bibinfo
  {author} {\bibfnamefont {H.}~\bibnamefont {Lai}}, \bibinfo {author}
  {\bibfnamefont {P.~M.}\ \bibnamefont {Nadolsky}},  \emph {et~al.},\
  }\href@noop {} {\bibfield  {journal} {\bibinfo  {journal} {JHEP}\ }\textbf
  {\bibinfo {volume} {0207}},\ \bibinfo {pages} {012} (\bibinfo {year}
  {2002})},\ \Eprint {http://arxiv.org/abs/hep-ph/0201195}
  {arXiv:hep-ph/0201195 [hep-ph]} \BibitemShut {NoStop}%
\bibitem [{\citenamefont {Kauer}(2007)}]{Kauer2007413}%
  \BibitemOpen
  \bibfield  {author} {\bibinfo {author} {\bibfnamefont {N.}~\bibnamefont
  {Kauer}},\ }\href {\doibase http://dx.doi.org/10.1016/j.physletb.2007.04.036}
  {\bibfield  {journal} {\bibinfo  {journal} {Physics Letters B}\ }\textbf
  {\bibinfo {volume} {649}},\ \bibinfo {pages} {413 } (\bibinfo {year}
  {2007})}\BibitemShut {NoStop}%
\bibitem [{\citenamefont {Barr}\ and\ \citenamefont
  {Lester}(2010)}]{Barr:2010zj}%
  \BibitemOpen
  \bibfield  {author} {\bibinfo {author} {\bibfnamefont {A.~J.}\ \bibnamefont
  {Barr}}\ and\ \bibinfo {author} {\bibfnamefont {C.~G.}\ \bibnamefont
  {Lester}},\ }\href {\doibase 10.1088/0954-3899/37/12/123001} {\bibfield
  {journal} {\bibinfo  {journal} {J.Phys.}\ }\textbf {\bibinfo {volume}
  {G37}},\ \bibinfo {pages} {123001} (\bibinfo {year} {2010})},\ \Eprint
  {http://arxiv.org/abs/1004.2732} {arXiv:1004.2732 [hep-ph]} \BibitemShut
  {NoStop}%
\bibitem [{\citenamefont {de~Favereau}\ \emph {et~al.}(2014)\citenamefont
  {de~Favereau} \emph {et~al.}}]{deFavereau:2013fsa}%
  \BibitemOpen
  \bibfield  {author} {\bibinfo {author} {\bibfnamefont {J.}~\bibnamefont
  {de~Favereau}} \emph {et~al.} (\bibinfo {collaboration} {DELPHES 3}),\ }\href
  {\doibase 10.1007/JHEP02(2014)057} {\bibfield  {journal} {\bibinfo  {journal}
  {JHEP}\ }\textbf {\bibinfo {volume} {1402}},\ \bibinfo {pages} {057}
  (\bibinfo {year} {2014})},\ \Eprint {http://arxiv.org/abs/1307.6346}
  {arXiv:1307.6346 [hep-ex]} \BibitemShut {NoStop}%
\end{thebibliography}%

\end{document}